\documentclass[a4paper,11pt]{article}
\pdfoutput=1
\usepackage{jheppub}

\usepackage{amsmath,amssymb,amsfonts,cases}
\usepackage{amsthm}
\usepackage{graphicx,caption,subcaption}
\usepackage{dsfont}
\usepackage{slashed}
\usepackage{wrapfig}
\usepackage{mathtools}
\usepackage{tikz}

\usepackage{tikz-cd}
\usepackage{physics}
\usepackage{hyperref}
\usepackage{simpler-wick}
\usepackage[inline]{enumitem}
\usepackage{mhchem}

\def\be{\begin{equation}}
\def\ee{\end{equation}}

\def\bal{\begin{equation}\begin{aligned}}
\def\eal{\end{aligned}\end{equation}}

\def\comment#1{}

\title{\LARGE An Atlas for 3d Conformal Field Theories with a U(1) Global Symmetry}
\author{Samuel Bartlett-Tisdall,}
\author{Christopher P.\ Herzog,}
\author{Vladimir Schaub}
\affiliation{Department of Mathematics, King's College London, \\  Strand, London, WC2R 2LS, UK}

\emailAdd{samuel.c.bartlett-tisdall@kcl.ac.uk}
\emailAdd{christopher.herzog@kcl.ac.uk}
\emailAdd{vladimir.schaub@kcl.ac.uk}

\abstract{ 
We present a collection of numerical bootstrap computations for 3d CFTs with a U(1) global symmetry. We test the accuracy of our method and fix conventions through a computation of bounds on the OPE coefficients for low-lying operators in the free fermion, free scalar, and generalised free vector field theories. We then compute new OPE bounds for scalar operators in the Gross-Neveu-Yukawa model, $O(2)$ model, and large $N$ limit of the $O(N)$ model. 
Additionally, we present a number of exclusion plots for such 3d CFTs. In particular, we look at the space of even and odd parity scalar operators in the low-lying spectrum that are compatible with crossing symmetry. As well as recovering the known theories, there are some kinks that indicate new unknown theories.
}

\makeatletter
\def\@fpheader{\vspace{0cm}}
\makeatother

\begin{document}
\maketitle

\section{Introduction}

%\sam{Missing some general motivation/context here}
%\chris{working on it}

In previous work  \cite{Bartlett-Tisdall:2023ghh}, we performed a numerical bootstrap of boundary QED (bQED). The numerical setup we considered can, in fact, be applied more generally to any 3d conformal field theory (CFT) with a U(1) global symmetry. In this paper, we use this setup to investigate the CFT data of such theories.  

Consider a 3d CFT with a U(1) global symmetry, whose conserved current is $J$. In this work, we bootstrap the correlation function of four such currents, $\langle JJJJ \rangle$, to obtain a number of bounds for OPE coefficients and scaling dimensions of such theories. 
Our effort here builds on a number of previous results in addition to ref.\ \cite{Bartlett-Tisdall:2023ghh}.
Our work can be considered a more or less direct extension and elaboration of ref.\ \cite{Dymarsky:2017xzb} from 2017.  Also of note are the more recent \cite{He:2023ewx} where the current bootstrap was extended to theories with a larger, nonabelian global symmetry group and \cite{Reehorst:2019pzi} where a mixed system involving currents and scalars was considered.  Our work makes use of some of the technical advancements of \cite{He:2023ewx}  in combination with the {\tt blocks$\_$3d} software \cite{Erramilli:2020rlr}, but otherwise does not look beyond U(1) symmetry and the four current correlation function.  

%
%Generally, correlation functions of four spin 1 operators will involve a large number of different structures, making the bootstrap problem not only complicated, but computationally expensive. However, we are able to use both the Bose symmetry and the conservation of $J$ to reduce the problem significantly.

We begin in section \ref{sec: setup} with a review of our setup in \cite{Bartlett-Tisdall:2023ghh}, and discuss details of the bootstrap. In particular, we describe how Bose symmetry and conservation reduce the problem, and outline the precise bootstrap problem we are solving. In appendix \ref{sec: theories}, we set conventions for the key theories that we investigate here. These are the free theories - the free scalar, free fermion, and generalised free vector field (GFVF) -- as well as the scalar $O(N)$ model and the Gross-Neveu-Yukawa (GNY) model. 
In a further appendix \ref{sec: numerics}, we provide the details of our numerical setup.
%We provide a brief review of the relevant details of this model.

In section \ref{sec: plots}, we present our results. We begin by looking at operator product expansion (OPE) coefficient bounds for scalar operators in the free fermion, free scalar, and GFVF theories. These results are presented in Table \ref{table:OPE}. To obtain this data, we input the spectrum of lowest-lying operators of the respective theories into our crossing equations.  In addition to fixing normalization conventions, this exercise gives some insight into the type of accuracy one might expect when these methods are applied to interacting theories where the spectrum is much less well known, which we do next 
%\sam{Sell main result of this more?}. 
In general, we find most bounds are within $10\%$ of the true value, with variations from under $1\%$ to nearly $30\%$.

Building on earlier numerical bootstrap work \cite{Erramilli:2022kgp, Kos:2015mba, Chester:2019ifh}, we are able to bound OPE coefficients of operators in the Gross-Neveu-Yukawa (GNY) and the $O(2)$ models. These models are some of the simplest 3d universality classes of fermions and bosons, respectively, that can be reached by an RG flow. Despite their simplicity, they describe the universality classes of several physical systems, providing tractable models that can be directly compared to experiment. For instance, the $N=8$ GNY universality class has been proposed to describe phase transitions in graphene \cite{Herbut_2006,Herbut_2009:1,Herbut_2009:2}. The $O(2)$ model is in the so-called XY universality class, describing phase transitions in (anti)ferromagnets, and the superfluid transition of \ce{^{4}He} \cite{Pelissetto_2002}. Due to the earlier work done with these models, in particular, isolating the lowest lying operators, we are able to investigate the higher field content. 
%\sam{Motivation - are these theories interesting, are there interesting 
%questions to answer using numerical conformal bootstrap? I also added some references}. 
We look at OPE coefficients of even and odd parity scalars in the GNY model. We consider $N=2,4,8$ fermions separately, again by inputting their lowest-lying even and odd parity operators, which were calculated in \cite{Erramilli:2022kgp}.\footnote{%
The calculation in ref.~\cite{Erramilli:2022kgp} uses the numerical conformal bootstrap. Their approach considers some bosonic and fermionic $O(N)$-symmetric fields, and bootstraps their mixed correlation functions, whilst our approach considers correlators of the associated $U(1)$ symmetry current.}
 %\sam{What was the method?}. 
 The results are shown in table \ref{table: GNY}, figure \ref{fig: GNY1}, and figure \ref{fig: GNY2}. We also look at the OPE coefficients for odd parity scalar operators in the O(2) model and the large $N$ limit of the $O(N)$ model. These results additionally give us upper bounds on the odd gaps, which are $\Delta_- < 5.71$ for the $O(N)$ model, and $\Delta_- < 7.57$ for the $O(2)$ model. The OPE bounds are shown in figure \ref{fig: O(N)}, and the $O(2)$ bound may agree with that found in \cite{Reehorst:2019pzi} although we are unsure of their conventions.
 %\sam{Are we still unsure?}

Lastly, we investigate a number of exclusion regions for scaling dimensions of even and odd scalar operators in the space of 3d CFTs with a U(1) symmetry. This is done for an even and odd scalar, two even scalars, and two odd scalars separately. We recover features associated to the known theories, but additionally we find behaviour indicating some unknown theories.

\section{Setup}\label{sec: setup}

\subsection{Theoretical Picture}

Consider any 3d CFT possessing a U(1) current $J$. The general bootstrap problem we are solving is to use crossing symmetry of the four current correlation function to constrain the space of such CFTs. 
Several examples of such theories are known, which focus our investigation, but our analysis is general and not based on a particular
Lagrangian starting point. Instead, we will make assumptions about the operator spectrum that are consistent with known examples to try to learn additional information about operator scaling dimensions and OPE coefficients for these examples.  
Within the framework of these assumptions, we also look for interesting features in our exclusion plots which may be indicative of yet unknown 3d CFTs.

The examples we focus on are the free scalar, free fermion, generalized free vector field, GNY model, and the $O(N)$ model.  Some conventions for these systems are gathered in appendix \ref{sec: theories}.
The first three examples are of course well understood, and our purpose in looking at free theories is not to derive new results but to validate the method and fix conventions for the results we derive for the GNY and $O(N)$ models.  
By inputting the first few low lying operators into the four current bootstrap, we are able to put relatively sharp bounds on the OPE coefficients, in some cases within a few percent of the analytically known answers (see Table \ref{table:OPE}).
With these results in hand, we analyze the OPE coefficients for the GNY and $O(N)$ models in a similar way.  Our bounds on the interacting theories
are unfortunately much less tight because we know much less about the low lying operator spectrum.  In the free case, we know the whole operator spectrum of which we are content to input information into the bootstrap about a half dozen low lying operators, while in the GNY and $O(2)$ case, we have information about just a couple of these low lying operators.

%
%We now outline a selection of such theories that we investigated. In particular, 
%we are interested in the low-lying scalar spectrum, since it is with that we are able to specify a theory using the numerical bootstrap. 

In the following, we denote scalar operators by $\mathcal{O}_{\Delta, p}$, where $\Delta$ is its conformal dimension, and $p =\pm 1$ is its parity. In the free theories, we are also able to analytically calculate three-point function coefficients $\gamma_{JJ\mathcal{O}}$ of correlators $\langle J J \mathcal{O} \rangle$. To compare to the bootstrap calculations, we define a normalized three-point coefficient 
\begin{equation}
    \lambda^2_{JJ\mathcal{O}} \equiv \frac{\gamma^2_{JJ \mathcal{O}}}{c_{JJ}^2c_{\mathcal{O} \mathcal{O}}},
\end{equation}
% \chris{fixed}
where $c_{JJ}$ and $c_{\mathcal{O} \mathcal{O}}$ denote normalization of the corresponding two-point functions. Note that $J(x, z) = J^\mu(x) z_\mu$ for some polarization vector $z$. These analytic values provide a way to verify our bootstrap.
The precise normalization is dependent on definitions of tensor structures used in the construction of $\langle JJO \rangle$ and $\langle JJ \rangle$, to which we refer the reader to \cite{Bartlett-Tisdall:2023ghh}.

\subsection{Numerical Setup}

The numerical problem that we solve in this work is the bootstrap of the correlation function
\begin{equation}
    \langle J(x_1) J(x_2) J(x_3) J(x_4) \rangle.
\end{equation}
We give a description of how this is done, with technical details in appendix \ref{sec: numerics}. The current $J$ is a spin 1 operator, and so there are initially 81 structures to include in our crossing equations, which we reduce to 41 by restricting to the parity preserving theory. Due to Bose symmetry, there is an additional $\mathbb{Z}_2 \times \mathbb{Z}_2$ permutation symmetry that also leaves the conformal cross ratios invariant. This allows us to reduce the number of structures from 41 to 17. Lastly, we use the conservation of $J$ to reduce the number of crossing equations that are undetermined to be 5, and take the schematic form
\begin{eqnarray}
  0 &=& \sum_{\Delta} \gamma_{0,\Delta} V_{0, \Delta} \gamma_{0,\Delta} +  \sum_{\Delta }  
 \widetilde \gamma_{0,\Delta} \widetilde V_{0, \Delta} \widetilde \gamma_{0,\Delta} \\
  &&
+   \sum_{\ell = 2}^\infty \sum_{\Delta }  \widetilde \gamma_{\ell, \Delta} \widetilde V_{\ell, \Delta} 
\widetilde \gamma_{\ell,\Delta} +
\sum_{\ell \in 2 {\mathbb Z}^+} \sum_\Delta ( \gamma^{1}_{\ell, \Delta}  \; \gamma^{2}_{\ell, \Delta} ) V_{\ell, \Delta} \left(
 \begin{array}{c}
  \gamma^{1}_{\ell, \Delta}  \\
  \gamma^{2}_{\ell, \Delta}  
 \end{array}
 \right) \ .
  \nonumber
 \end{eqnarray}
Here, $\gamma_{\ell, \Delta}$ and $V_{\ell, \Delta}$ indicate the OPE coefficients and crossing vector for a given spin $\ell$ and scaling dimension $\Delta$, respectively. A tilde indicates odd parity, whilst its absence indicates even parity. The tensor structure of this crossing equation is explained in appendix \ref{sec: numerics}.

To investigate a given theory, we insert the lowest-lying operators in the spectrum. That is, we evaluate the relevant crossing vectors at the scaling dimensions corresponding to the operator in question, and input them as 
% singletons \sam{non-standard, 'Isolated operators'}.
isolated operators. We then additionally include some gap assumptions for the remaining crossing vectors, before running the bootstrap. This can be done to produce both exclusion regions in the space of scaling dimensions, and to bound OPE coefficients of particular operators. We do both of these and present our results in the following section.

\section{Results}\label{sec: plots}

We present our results. In general, our plots are compared with different parameters. One parameter that is always compared is $\Lambda$, the derivative order up to which the bootstrap satisfies. We compare a smaller value of $\Lambda = 17$ or $18$ to a larger one, $\Lambda = 25$. This gives us an indication of the convergence of the bounds.

\subsection{OPE Peninsulas}

\begin{figure}[h]
\begin{center}
\includegraphics[width=3in]{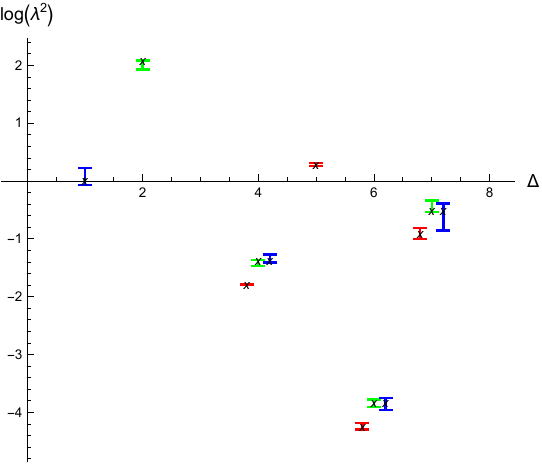}
\end{center}
\caption{\label{fig: OPE}
OPE coefficients for the free scalar (blue), free fermion (green), and generalized free vector field (red) along with upper and lower bounds computed from SDPB.  When more than one theory has an operator of a given dimension, the theories are displaced
slightly from each other to make the plot easier to read.  (The scaling dimensions are all integers for these theories.)
To try to pick out the free theories, the first few operators in their spectrum have been packed as isolated operators in SDPB.
Also, a gap of one has been introduced above the stress tensor for the free fermion and scalar.  For the GFVF, the spin two gap
is assumed to be $\Delta_2 > 4$.  The runs here had $\Lambda = 25$.  The $x$ marks the actual OPE coefficient and the error 
bars give the upper and lower bounds from SDPB.  The plot is a graphical representation of the data in the table \ref{table:OPE}.
}
\end{figure}

We begin by bounding the OPE coefficients for operators in the free scalar, free fermion, and GFVF, as discussed in section \ref{sec: setup}. The data is given in table \ref{table:OPE}, or graphically in figure \ref{fig: OPE}. We see that the actual values fall between the upper and lower bounds in every case.  In some cases, for example the $4_0^+$ GFVF scalar, the bounds are quite tight, within a few percent of the actual value.  More to the point, however, we get a sense of how well this method may work in cases where we don't already know the answer -- the GNY model and the $O(N)$ model.
(It would be interesting to explore additional constraints on this data, for example from integrability like in \cite{Cavaglia:2022qpg}, but 
we leave such endeavors for the future.)
 We also are able to fix and be certain of the normalization.

An interesting feature of the free theories is that there is a dimension five odd parity scalar for the GFVF but not for the free fermion or free scalar \cite{Dymarsky:2017xzb}.  The absence of this dimension five operator is an artifact of having only $U(1)$ global symmetry.  Theories with additional free scalars or fermions that enhance the global symmetry to $O(N)$ where $N>3$ will have such a dimension five operator. For more details, see the discussion at the end of appendix \ref{sec: theories}.
% (See the discussion at the end of appendix \ref{sec: theories} for more details.) 
The bounds we see for the interacting theories below echo this fact in an interesting way, as discussed in our analysis of the GNY model below. 
%\sam{Wording might be a bit confusing here}

% \chris{wondering if we should redo this but just including the lowest operators, to have something more comparable to what we do in the GNY and O(N) cases.}

\begin{table}[h]
{\small
\begin{tabular}{|c|c|c|c|c|c|c|c|c|}
\hline
& & & \multicolumn{2}{|c|}{gaps} & \multicolumn{3}{|c|}{OPE coefficient} \\
\cline{4-8}
theory & $\Delta^{\rm parity}_{\rm spin}$ & isolated operators &  even & odd &  lower bound & upper bound & actual \\
\hline
free  & $2^-_0$ & $4^+_0$,  $7^-_0$, $3^\pm_2$ & 6 & 9  & 6.90 & 8.05 & 8 \\ 
\cline{2-8}
fermion  & $7^-_0$ & $2^-_0$, $4^+_0$, $3^\pm_2$  & 6 & 9  & 0.584 & 0.715  & $\frac{3}{5}$  \\
\cline{2-8}
& $4^+_0$ & $2^-_0$, $7^-_0$, $3^\pm_2$ & 6 & 9  & 0.231 & 0.254 & $\frac{1}{4}$ \\
\cline{2-8}
& $6^+_0$ & $2^-_0$, $4^+_0$, $7^-_0$, $3^\pm_2$ & 8 & 9  & 0.0201 & 0.0229 & $\frac{3}{140} \approx 0.021$ \\
\hline
free &
$7^-_0 $ & $1_0^+$, $4_0^+$, $3^\pm_2$ & 6 & 9  & 0.425 & 0.677 & $\frac{3}{5}$ \\
 \cline{2-8}
scalar  & $1^+_0$  & $4_0^+$, $7_0^-$, $3^\pm_2$ & 6 & 9  & 0.938 & 1.25 & 1 \\
\cline{2-8}
& $4^+_0$  & $1_0^+$, $7_0^-$, $3^\pm_2$ & 6 & 9  & 0.244 & 0.281 & $\frac{1}{4}$ \\
\cline{2-8}
& $6^+_0$ & $1_0^+$, $4_0^+$, $7_0^-$, $3^\pm_2$ & 8 & 9  & 0.0191 & 0.0234 & $\frac{3}{140} \approx 0.021$
\\
\hline
GFVF &
$5^-_0$ & $4_0^+$, $6_0^+$, $7_0^-$, $9_0^-$ & 8 & 11  & 1.294 & 1.362 & $\frac{4}{3}$ \\
\cline{2-8}
& $7^-_0$ & $4_0^+$, $6_0^+$, $5_0^-$, $9_0^-$ & 8 & 11  & 0.368 & 0.442 & $\frac{2}{5}$ \\
\cline{2-8}
& $4^+_0$ & $6_0^+$, $5_0^-$, $7_0^-$, $9_0^-$  &  8 & 11  & 0.165 & 0.168 & $\frac{1}{6}$ \\
\cline{2-8}
& $6^+_0$ & $4_0^+$, $5_0^-$, $7_0^-$, $9_0^-$ & 8 & 11  & 0.0136 & 0.0153 & $\frac{1}{70} \approx 0.014$\\
\hline
\end{tabular}
}
\caption{
Bounds on OPE coefficients for free theories. The isolated operators are those input into SDPB to specify the theory, along with the corresponding even and odd parity scalar gap assumptions in the next column. In all cases, the continuum part of the spin 2 blocks starts at $\Delta_2 = 4$. 
\label{table:OPE}
}
\end{table}

\begin{figure}[h]
\begin{center}
a) \includegraphics[width=2.5in]{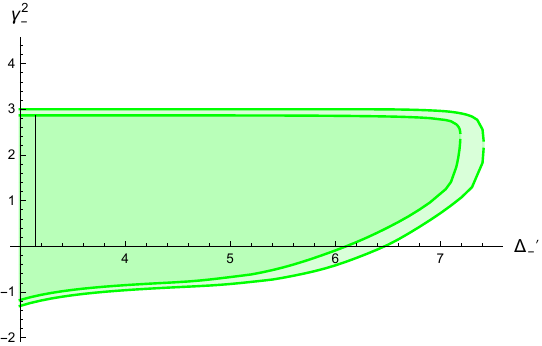}
b) \includegraphics[width=2.5in]{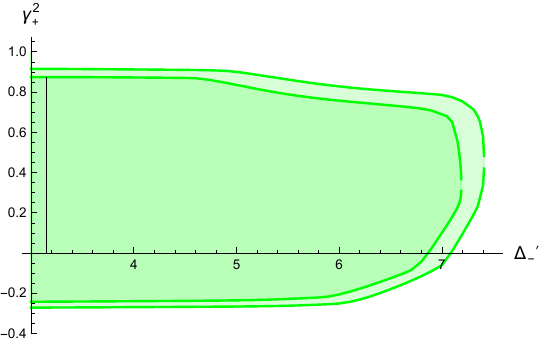}\\
c) \includegraphics[width=2.5in]{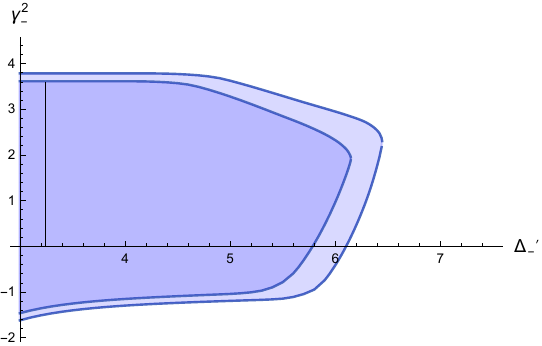}
d) \includegraphics[width=2.5in]{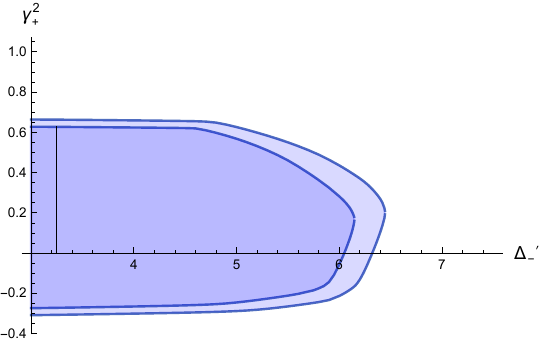}\\
e) \includegraphics[width=2.5in]{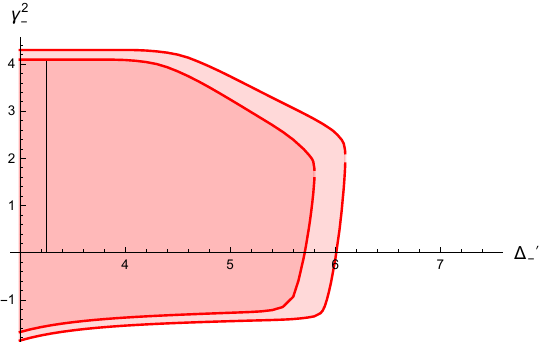}
f) \includegraphics[width=2.5in]{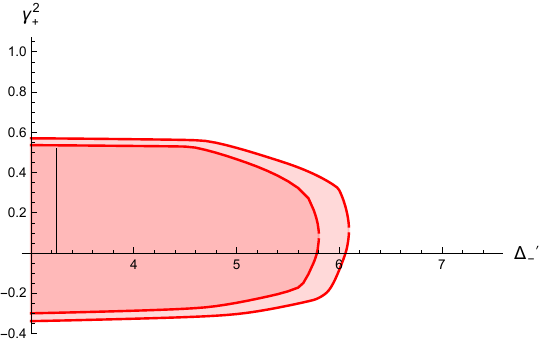}
\end{center}
\caption{\label{fig: GNY1}
% \sam{Mention GNY} 
Allowed region for the square of an OPE coefficient $\gamma^2_\pm$ vs.\ the gap $\Delta'_-$ 
to the second lowest odd scalar primary operator in the GNY model for different numbers of fermions.
For the plots on the left, the OPE coefficient is that of the lowest odd parity scalar primary.  For the plots on the right,
the OPE coefficient is the lowest even parity scalar primary.  In each case, two sets of curves are plotted.  The larger
allowed region comes from a run with $\Lambda =18$, the smaller with $\Lambda = 25$.  
For unitary theories, $\gamma^2$ should be positive and the region below the $x$-axis is excluded.
To bound the lowest dimension odd/even parity scalar OPE coefficient, the lowest dimension 
even/odd parity scalar has been packed as an isolated operator for input to SDPB.
We have assumed a gap $\Delta'_+>3$ above the lowest even parity scalar primary.  The odd gap assumption is the $x$-axis.
The top row is for the $N=2$ theory ($\Delta_-
 = 0.65$, $\Delta_+ = 1.725$), the middle row $N=4$ ($\Delta_-
= 0.7578$, $\Delta_+ = 1.899$), and the bottom row $N=8$ ($\Delta_-
= 0.8665$, $\Delta_+ = 2.002$).  
The black vertical lines indicate computed locations of the second lowest odd parity primary in these theories \cite{Mitchell:2024hix}.
% \sam{$N$ is fermion flavours? Also how to understand the region where $\gamma^2_\pm \neq 0$ (or anywhere to the right of the region)? Does it mean continuum of operators must exist in theory?? Or I suppose if you fix $\Delta_-$ and put in the operator, the picture to the right will change, since the x axis is just a lower bound. So it just means you need to have some operator in the region between $\gamma^2_\pm \neq 0$ and the end of the island...? Yes} 
%\sam{Also looks like by comparison with Petr's paper this is $\Delta_\pm - 0.5$} \sam{In the paper, $\sigma$ is parity odd, and $\epsilon$ is parity even. If that is the case, it seems like the values typed here are opposite from their tables, i.e. we have said $\Delta_+ = \Delta_\sigma$ and $\Delta_- = \Delta_\epsilon$, when it should be the other way round. Is this just a typo here, a possible problem with the bootstrap, or am I misunderstanding?}
%\chris{fixed}
}
\end{figure}

\begin{figure}[h]
\begin{center}
a) \includegraphics[width=2.5in]{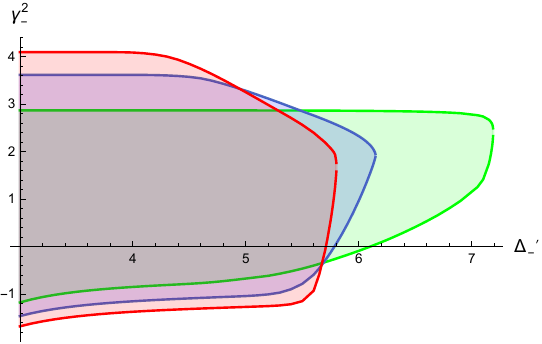}
b) \includegraphics[width=2.5in]{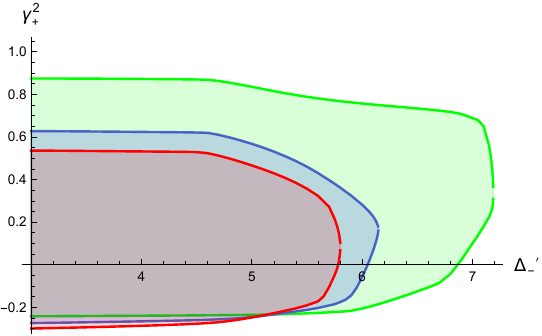}
\end{center}
\caption{\label{fig: GNY2}
Allowed region for the square of an OPE coefficient $\gamma^2_\pm$ vs.\ the gap $\Delta'_-$ 
to the second lowest odd scalar primary operator. Here we have superposed the different theories $N=2$, 4, and 8 for the $\Lambda = 25$ run.
}
\end{figure}

Next, we look at the GNY model for $N=2, 4$ and $8$.  We restrict to these theories by inputting the lowest even and odd parity scalars as isolated operators in our bootstrap.  The dimensions of these operators were calculated in ref.\ \cite{Erramilli:2022kgp} also using the numerical conformal bootstrap, but for a different system of correlation functions.
% \sam{Comment on their setup, is it different?}. 
Their dimensions are $\Delta_+ = 0.65, \Delta_- = 1.725$ for the $N=2$ theory, $\Delta_+ = 0.7578, \Delta_- = 1.899$ for the $N=4$ theory, and $\Delta_+ = 0.8665, \Delta_- = 2.002$ for the $N=8$ theory. We then bound the OPE coefficient of both these even and odd scalars as a function of the dimension $\Delta'_-$ of the second lowest odd scalar primary. 
%
%
%\sam{Clarify what $\Delta'_-$ is - in the caption, Chris describes it as the gap to the second lowest odd scalar primary operator. 
%I'm assuming by gap this means from 0, rather than from the first lowest odd scalar primary?} \chris{yes}. 
%
These exclusion plots are shown in figure \ref{fig: GNY1}, where $\Lambda$ is compared, and figure \ref{fig: GNY2}, where the different values of $N$ are compared. The maximum values are summarised in table \ref{table: GNY}.

\begin{table}[h]
\begin{center}
\begin{tabular}{|c|c|c|c|}
\hline
theory & max $\Delta_-'$ & max $\gamma_+^2$ & max $\gamma_-^2$ \\
\hline
$N=2$ & 7.19 & 0.875 & 2.87 \\
$N=4$ & 6.15 & 0.628 & 3.62 \\
$N=8$ & 5.81 & 0.537 & 4.10 \\
\hline
\end{tabular}
\end{center}
\caption{
\label{table: GNY}
%\sam{Actually, what is this table? I don't see what the numbers are matching (except the max $\Delta_-'$, which matches the plots for $\Lambda = 25$ I think)}
% \sam{The table is correct, the plot should be $2\gamma^+_\pm$ on the axis}}
% \chris{fixed}
Here, $\operatorname{max}\; \Delta_-'$ is the maximum value of the scaling dimension of the second odd parity scalar in the corresponding GNY model.  The lowest odd parity scalar is an input.  These values can be eyeballed from figures \ref{fig: GNY1} and \ref{fig: GNY2} by looking at how far to the right the colored regions extend.  Also $\operatorname{max} \gamma_\pm^2$ is the maximum allowed OPE coefficient of the lowest odd or even parity scalar.  These values can also be eyeballed from figures \ref{fig: GNY1} and \ref{fig: GNY2} by looking at how high the colored regions extend.
}
\end{table}

An interesting feature of these bounds is that $\Delta_-'$ decreases as $N$ increases.  The $N=2$ result, with its smaller global symmetry
and bound $\Delta_-' < 7.19$ is consistent with the free theory result with $O(2)$ global symmetry which has a dimension 7 odd parity scalar but lacks the dimension 5 operator.  The $N=4$ and 8 results, with their larger global symmetry, must have an odd parity scalar with dimension less than 7, possibly close to 5, again in alignment with what happens for the free theories.  

Recent results \cite{Mitchell:2024hix} show that this bound on $\Delta_-'$ that we find is not tight and may be a kind of echo of the free theory result rather than telling us about the operator spectrum of the GNY model.  The authors of ref.\ \cite{Mitchell:2024hix}  find from bootstrapping a mixed fermion-scalar system that
 $\Delta_-' <4$, well below $\Delta =5$ or 7 in all cases.  
(Indeed using a Pad\'e approximation, they find the more precise values 
$\Delta_-' = 3.1487$, 3.2455, and 3.24739 for the $N=2$, 4, and 8 GNY models respectively.)
As the rigorous bootstrap bounds they find on the even parity irrelevant scalars are somewhat broad, for example $3.169 < \Delta_+' < 3.835$ in the $N=2$ case, we have decided not to input
them as isolated operators in our bootstrap.  
% \chris{wondering if we should rerun for $\Delta_+' > 3.169, 3.137, 3.167$ respectively - expensive :( }
Note the upper bound on $\gamma_\pm^2$ is relatively flat as a function of small values of $\Delta_-$.

The next plot looks at the OPE coefficient of the lowest even parity scalar in the $O(2)$ model and $O(N)$ model at large $N$. We look at how the bounds scale against the odd parity scalar gap. We make some assumptions about the dimension of the even scalar; for the $O(2)$ model, we assume the gap to the next even parity scalar is $\Delta_+' > 3.7$, whilst for the $O(N)$ model, we assume it is $\Delta_+ > 4$. Our results are possibly consistent with \cite{Reehorst:2019pzi} although we are uncertain of their conventions. 
%\sam{Are we still unsure?}

\begin{figure}
\begin{center}
\includegraphics[width=3in]{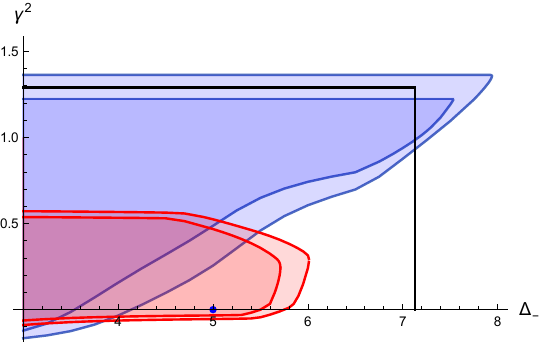}
\end{center}
\caption{\label{fig: O(N)}
Bounds on the OPE coefficient of the lowest even parity scalar vs.\ the odd parity scalar gap. 
The blue regions correspond to the O(2) model which has a lowest even parity scalar with $\Delta_+ = 1.51136$.  
It is also assumed that the next even parity scalar has a dimension at least $\Delta_+' > 3.7$.  
The stress tensor is inserted as an isolated operator with an aggressive gap of 1.6 to the next spin 2 operator.
The red regions correspond to the large $N$ limit of the $O(N)$ model, which has a lowest even parity scalar with
$\Delta_+ = 2$ and a gap to the next scalar of $\Delta_+' > 4$.  Here the gap between the stress tensor and the next spin 2 is assumed to be only one. 
The runs in both cases were with $\Lambda = 25$ (inner region) or $\Lambda = 18$ (outer region).  
The nearly horizontal line at the top of the blue region implies $\gamma^2  < 1.225$ for the $\Lambda = 25$ run.  
The largest value of $\gamma^2$ allowed in the large $N$ limit for the $\Lambda = 25$ run is 
0.538.
We can also conclude the odd gap $\Delta_-<5.71$ for the $O(N)$ model at large $N$ and $\Delta_-<7.57$ for the
$O(2)$ model.
Because the large $N$ model is close to non-interacting up to $1/N$ corrections, the OPE coefficient should vanish in this case (the blue dot).  The black horizontal line is the predicted value of the OPE coefficient
from \cite{Reehorst:2019pzi} although it might be half this.  
[We would like to thank M.~Reehorst for valiant attempts to match conventions with ours.]
%\sam{Are we still worried about the half? 
%Also is the square bracket notation standard?}
The black vertical line is the prediction from ref.\ \cite{Reehorst:2019pzi} of an upper bound on the dimension of the odd parity scalar 
$\Delta_- < 7.13$. }
\end{figure}

\clearpage
\subsection{Exclusion Islands and Lakes}

In this section, we look at allowed theories in the space of both even and odd scalar dimensions. These plots are produced by inserting two isolated scalar operators at particular dimensions, along with appropriate gap assumptions. Then either such a theory is excluded, or else it is allowed.

\begin{figure}[hbt!]
\begin{center}
 \includegraphics[width=2.7in]{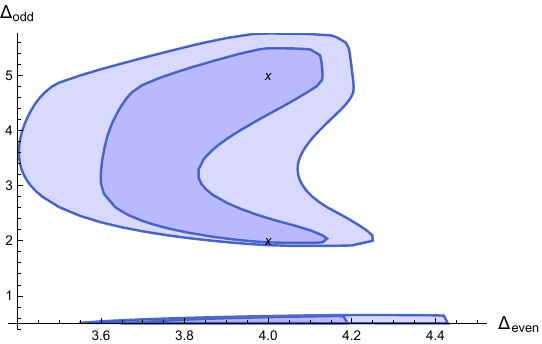}
 \includegraphics[width=2.7in]{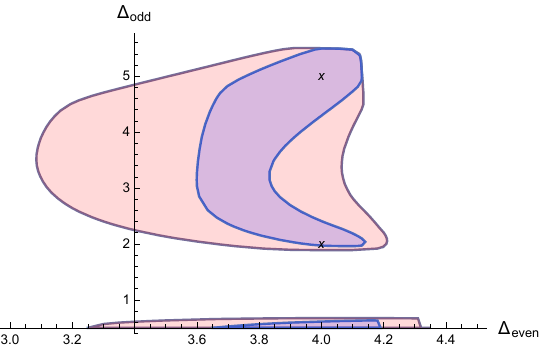}
\end{center}
\caption{\label{fig: even odd}
The allowed region for a conformal field theory assuming a single even parity scalar and a single odd parity scalar
below a threshold.  In both cases, the stress tensor is packed as an isolated operator with a gap of one to the next spin 2 operator.
On the left, we are comparing the effect of $\Lambda$ on the island.  The smaller island has a run with $\Lambda =25$ and the larger
with $\Lambda = 17$.  For both runs, the gap to the next scalar operators are $\Delta_{+}'>6$ and $\Delta_{-}' > 7$.  On the right, we are comparing the effect of the scalar gap.  The smaller island assumes $\Delta_{+}'>6$ and $\Delta_{-}' > 7$, whereas the larger island assumes $\Delta_{+}'>5.5$ and $\Delta_{-}' > 6.5$.  In both cases, the $x$'s mark the location of the free fermion and GFVF.  
}
\end{figure}

The first plot, figure \ref{fig: even odd}, shows the allowed region for an even scalar and an odd scalar. Note that the stress tensor is packed as an isolated operator, and then a gap of one is included to the next spin 2 operator. Importantly, the region includes the free fermion and GFVF. As $\Lambda$ increases, we can see the region between these two regions shrinks, though maintains the same curved shape. This could indicate some theory or even curve of theories connecting the two isolated free theories that would become clear in the large $\Lambda$ limit.

\begin{figure}
\begin{center}
 \includegraphics[width=2.7in]{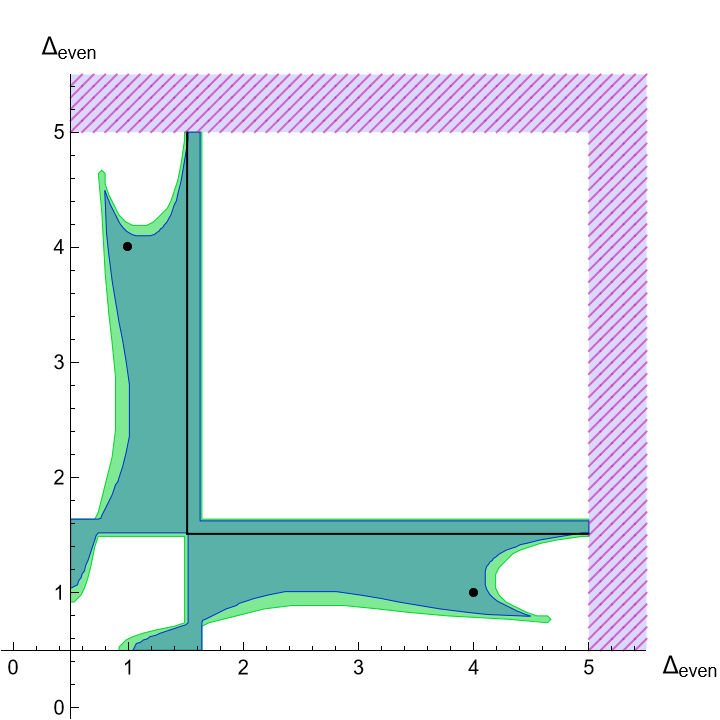}
 \includegraphics[width=2.7in]{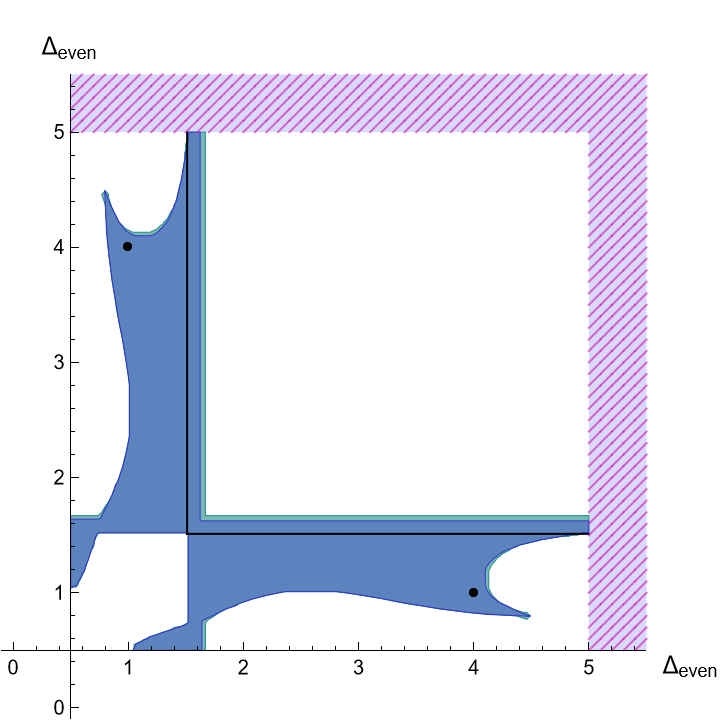}
\end{center}
\caption{\label{fig: even even}
The allowed region for a conformal field theory assuming two even parity scalars below a threshold. For all runs, the gap to the next even scalar operators is $\Delta_{+}''>5$, which is where the shaded region starts along the top and right edges of the plots. The line at these cutoffs can be understood as a single scalar inserted below the threshold. In both cases, the stress tensor is packed as an isolated operator with a gap of one to the next spin 2 operator. The dot indicates the free scalar, whilst the line indicates the $O(2)$ model.
On the left, we are comparing the effect of $\Lambda$ on the region. The smaller region has a run with $\Lambda =25$ and the larger with $\Lambda = 18$. For both runs, the gap to the next odd scalar operators is $\Delta_{-}' > 7$.
On the right, we are comparing the effect of the odd scalar gap at a fixed value of $\Lambda = 25$. The smaller region assumes $\Delta_{-}' > 7$, whereas the larger region assumes $\Delta_{-}' > 6.5$.
Note that the rigid lines are a feature of the plot, rather than due to a lack of precision. We have also used different colours to make different outlines clearer, as they are close together. Note that smaller regions are in the foreground, lying on top of the larger regions.
% \sam{Remake with white background, make connection to bulk region clearer, thicker lines to match Chris', add free scalar marker}
}
\end{figure}

The next plot, figure \ref{fig: even even}, shows the allowed region for two even parity scalars. The dot indicates the free scalar, whilst the line indicates the $O(2)$ model. The empty square on the upper right hand side indicates that a low-lying even scalar is required, with a bound possibly set by the $O(2)$ model in the large $\Lambda$ limit. Similarly, the exclusion region in the bottom left indicates that there cannot be two low-dimension scalars, except on the small legs that connect the island to the unitarity bound. Note the leg shooting off near the free scalar. In particular, the tip of the leg stays far from the free theory, even as $\Lambda$ increases. This pointy tip could indicate the existence of an unknown theory.
% \chris{whether to run spectrum on this point?}
% in that region keeping the leg away from the free scalar even in the large $\Lambda$ limit.

\begin{figure}
\begin{center}
 \includegraphics[width=2.7in]{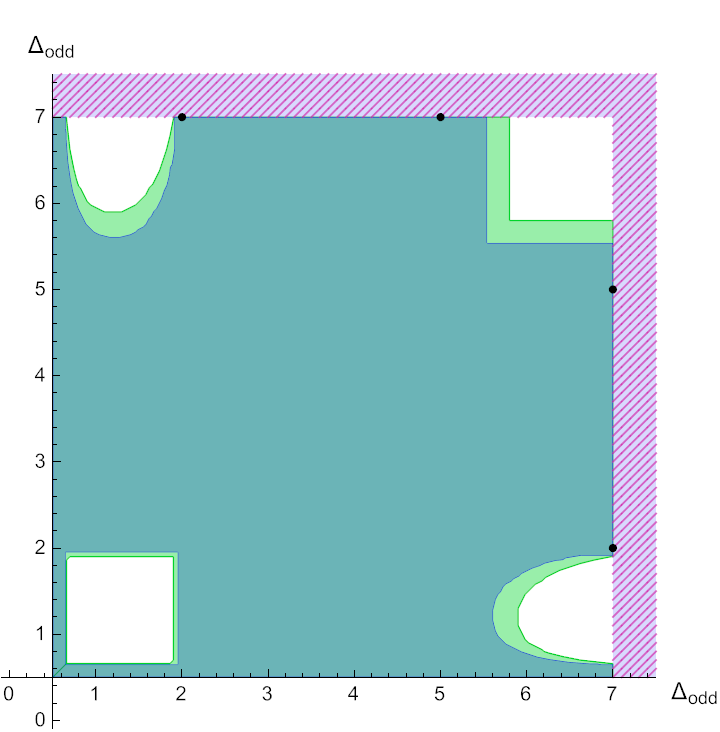}
 \includegraphics[width=2.7in]{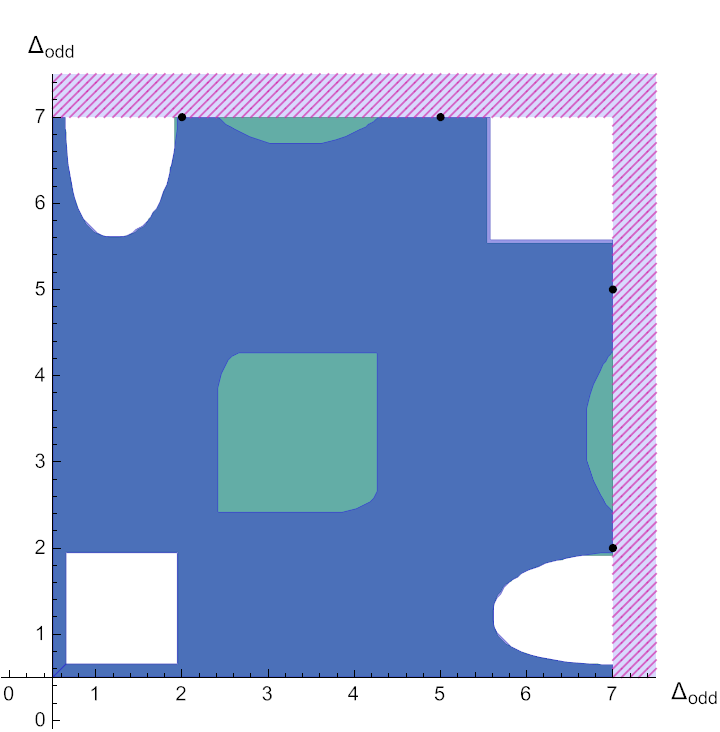}
\end{center}
\caption{\label{fig: odd odd}
The allowed region for a conformal field theory assuming two odd parity scalars below a threshold. For all runs, the gap to the next odd scalar operator is $\Delta_{-}''>7$, which is where the shaded region starts along the top and right edges of the plots. The line at these cutoffs can be understood as a single scalar inserted below the threshold. In both cases, the stress tensor is packed as an isolated operator with a gap of one to the next spin 2 operator.
On the left, we are comparing the effect of $\Lambda$ on the region. The smaller region has a run with $\Lambda =25$ and the larger with $\Lambda = 18$. For both runs, the gap to the first even scalar operator is $\Delta_{+} > 3.5$.
On the right, we are comparing the effect of the even scalar gap at a fixed value of $\Lambda = 25$. The smaller region (the blue, to be clear) assumes $\Delta_{+} > 4$, whereas the larger region assumes $\Delta_{+} > 3.5$. 
% \chris{typo on subscript?}
The dots mark the free fermion and GFVF. Again, note that the rigid lines are a feature of the plot, rather than due to a lack of precision. Similarly, the smaller regions overlay the larger regions again.
% \sam{thicker lines to match Chris', make connection to bulk region clearer}
}
\end{figure}

The final exclusion region in figure \ref{fig: odd odd} is that for two odd scalars. Like figure \ref{fig: even even}, there is an empty square in the upper right corner, indicating that the two odd scalars cannot both have a large dimension. Although the GFVF is still a distance from this cutoff, given the rapid change of this region with $\Lambda$, it is possible that the GFVF could be the upper bound. The bay next to the free fermion is also interesting. Whilst on the right hand side (that is, when looking at the bay in the top left corner) it appears bounded by the free fermion, it also appears to be bounded on the left hand side, perhaps indicating another theory here. Similarly, there is an exclusion region in the bottom left corner, which appears strictly bounded in both dimensions. Again, this could indicate some theories that keep these bounds here even in the large $\Lambda$ limit. 

\section{Discussion}

We now discuss some of our results, and some potential future directions. Firstly, consider our results for the GNY model, as shown in figures \ref{fig: GNY1} and \ref{fig: GNY2}, and table \ref{table: GNY}. 
%
%As we mentioned, when compared to the recent results in \cite{Mitchell:2024hix}, our weak bounds tell us little about the GNY model. 
%
Optimistically, the OPE coefficients $\gamma_+$ and $\gamma_-$ may saturate the upper bounds we found in Figures \ref{fig: GNY1} and  \ref{fig: GNY2}.  The behavior of these bounds as a function of $\Delta_-'$ is relatively flat, leading potentially then to good predictions for these coefficients.  
The constraint on $\Delta'_-$ on the other hand looks relatively weak, especially compared with the results from ref.\ \cite{Mitchell:2024hix} where their Pad\'e approximates are the black vertical lines in these same figures.  
Instead, for us the maximal values of $\Delta'_-$ in table \ref{table: GNY} are more comparable to the dimension 5 and 7 operators in the free theories. It appears to be the free theory operators that extend the region where the OPE coefficients are non-zero. This extension could be a numerical artifact -- the truncated crossing equations are unable to distinguish these theories sufficiently -- or alternatively, it may indicate a broader relation between the free theory and GNY model. 

Regarding the $O(N)$ models, if the horizontal black line in Figure \ref{fig: O(N)} is indeed the properly normalized result from \cite{Reehorst:2019pzi}, then presumably there is some small fix either in the calculations here or the calculations in \cite{Reehorst:2019pzi} that would reconcile the two predictions.  Possibly the error in \cite{Reehorst:2019pzi} was misreported, or perhaps the failure to see dual-jumps in our system, described in appendix \ref{sec: numerics}, introduces a small amount of slop into our bounds.  Assuming the two results can be reconciled, the result in ref.\ \cite{Reehorst:2019pzi} then suggests the OPE coefficient $\gamma$ of the lowest even parity scalar for the $O(2)$ model may saturate our upper bound in Figure \ref{fig: O(N)}.  More exciting, the lowest odd parity operator may lie close to the cusp at $\Delta_- \sim 7.5$. 
Of course the amount the blue region shrinks on increasing $\Lambda$ from 25 and 18 casts some shade on these ideas, but it would be interesting to try to pin down the operator dimension $\Delta_-$ in the future and the OPE coefficient $\gamma$.

For the exclusion plots shown in figures \ref{fig: even odd}, \ref{fig: even even}, and \ref{fig: odd odd}, there is some potentially interesting behavior. In figure \ref{fig: even odd}, the free theories remain on the same island. It is unclear whether this connection would remain in the large $\Lambda$ limit, perhaps indicating some theories or curve of theories connecting the two. Alternatively, it may be possible to isolate the theories at higher order.

For the two even parity scalar exclusion plots in figure \ref{fig: even even}, the structure also hints at something new. The plots appear to converge well with $\Lambda$, and yet there remains a spike above the free scalar. It would be interesting if there was some isolated theory in that region, keeping the plot from pulling in to the free theory. We attempted to look at the operator spectrum in this region, using the \texttt{spectrum} routine packaged with SDPB \cite{sdpb,Landry:2019qug}, along with a manual inspection of the functionals. Our results were inconclusive. (As a test case, we also attempted the same analysis on the free scalar theory, which did not yield the expected spectrum either.)  We would like to try and explore this cusp further in the future.

A similar story holds for the two odd scalar exclusion plots in figure \ref{fig: odd odd}. Here, there are a few regions of interest. Firstly, the bay in the top left that is bounded on one side by the free fermion appears to be similarly bounded on the other side. Again, this appears to converge in $\Lambda$, perhaps indicating an isolated theory here. The square in the bottom left corner also displays interesting behavior, and we wonder if the sharp cutoffs have a stronger physical meaning.

%\sam{Unfortunately don't have much to say about the integrability side of things if there is anything there... }

% \sam{Some notes on producing the plots manually: from discussion with Marten, I understand that the output y.txt is related to the functional $\alpha$ by some kind of linear transformation, which somehow encodes the normalisation condition. This is invertible, but I'm not exactly sure how, Marten said might be able to send over a relevant function to do this. Then, it's simply a matter of implementing the contraction of $\alpha$ with the crossing vector, to get the polynomial. A log plot should make the zeroes clearer.}

\section*{Acknowledgments}

We would like to thank P.~Kravchuk, S.~Minwalla, M.~Reehorst, R.~Sinha, and A.~Stergiou for discussion.
This work was supported in part by was supported by a Wolfson Fellowship from the Royal Society and by the
U.K. Science and Technology Facilities Council Grant ST/P000258/1.
Computations were performed on the HPC CREATE KCL cluster, and COSMA8 at the University of Durham.

\clearpage

\appendix

\section{Conventions for 3D CFTs} \label{sec: theories}

\subsection*{Free Fermion}

The free fermion has the Green's function
\begin{equation}
    \langle \psi(x) \bar \psi(0) \rangle = \frac{\gamma_\mu x^\mu}{|x|^3} \ 
\end{equation}
constructed from two fermionic operators. The current in this theory is
\begin{equation}
    J^\mu = \bar{\psi} \gamma^\mu \psi \ ,
\end{equation}
with two-point function
\begin{equation}
    \langle J^\mu(x) J^\nu(0) \rangle = 2 I^{\mu\nu}(x) / |x|^4 .
\end{equation}
Here, we use 
\begin{equation}
    I^{\mu\nu}(x) = \delta^{\mu\nu} - \frac{2 x^\mu x^\nu}{|x|^2}.
\end{equation}
In three dimensions mass breaks parity, and so the lowest lying scalar is the parity odd $\mathcal{O}_{2,-} = \bar{\psi} \psi$. There is a parity even operator\footnote{A Fierz identity relates $( \bar{\psi} \psi)^2 \sim J^\mu J_\mu$.} $\mathcal{O}_{4, +} = ( \bar{\psi} \psi)^2$.
The relevant three-point function coefficients can be calculated to be
\begin{equation}
    \begin{array}{c|ccc}
    & c_{\mathcal{OO}} & \gamma_{JJ \mathcal{O}} & \lambda^2_{JJ \mathcal{O}} \\
    \hline
    {\mathcal O}_{4,+} & 4  & 2  & \frac{1}{4} \\
    {\mathcal O}_{2,-} & 2 & 8 &  8  \\
    \end{array}
\end{equation}
We did not compute the corresponding data for ${\mathcal O}_{6,+}$ and ${\mathcal O}_{7,-}$, but instead observe that the OPE coefficients are expected to be the same as in the free scalar model, looking at \cite{Dymarsky:2017xzb}.  

\subsection*{Free Scalar}

For the free scalar, we have the Green's function
\begin{equation}
   \langle \phi(x) \phi(0) \rangle = \frac{1}{|x|} \ . 
\end{equation}
The current is
\begin{equation}
    J_\mu = i (\phi \partial_\mu \bar \phi - \bar \phi \partial_\mu \phi)  \ ,
\end{equation}
and similarly to the free fermion,
\begin{equation}
    \langle J^\mu(x) J^\nu(0) \rangle = 2 I^{\mu\nu}(x) / |x|^4 .
\end{equation}
The lightest operators are 
\begin{eqnarray}
\mathcal{O}_{1, +} &=& | \phi |^2, \\
{\mathcal O}_{4,+} &=& (\partial_\mu \phi) (\partial^\mu \bar \phi) |\phi|^2 - \frac{1}{2} (\partial_\mu \phi)(\partial^\mu \phi) \bar \phi^2 - \frac{1}{2} (\partial_\mu \bar \phi) (\partial^\mu \bar \phi) \phi^2 \ , \\
{\mathcal O}_{6,+} &=&  
\frac{5}{2}\biggl( [ (\partial_\mu \phi) (\partial^\mu \bar \phi) ]^2 -  |  (\partial_\mu \phi) (\partial^\mu \phi)|^2 \biggr)
+ \frac{1}{6} (\partial_\mu \partial_\nu \phi) (\partial^\mu \partial^\nu \bar \phi) |\phi|^2 \nonumber \\
&&
 - \frac{1}{12} \biggl( (\partial_\mu \partial_\nu \phi) (\partial^\mu \partial^\nu \phi)  \bar \phi^2 
-(\partial_\mu \partial_\nu  \bar \phi) (\partial^\mu \partial^\nu \bar \phi)  \phi^2 \biggr)
\\ 
&& + (\partial_\mu \partial_\nu \phi) \biggl((\partial^\mu \bar \phi) (\partial^\nu \phi) \bar \phi
- (\partial^\mu \bar \phi) (\partial^\nu  \bar \phi) \phi \biggr) \nonumber \\
&&
+  (\partial_\mu \partial_\nu \bar \phi)  \biggl(  (\partial^\mu  \phi) (\partial^\nu \bar \phi)   \phi -
(\partial^\mu \phi) (\partial^\nu  \phi)  \bar \phi
 \biggr)
\ ,
\nonumber \\
{\mathcal O}_{7,-} &=& i \epsilon^{\mu\nu\lambda} \left[2 J_\mu (\partial^\rho \partial_\nu \phi )(\partial_\rho \partial_\lambda \bar \phi) 
+3(\partial^2 J_\mu)  ( \partial_\nu \phi )(\partial_\lambda \bar \phi) \right] .
\end{eqnarray}
We find the following three-point function coefficients:
\begin{equation}
    \begin{array}{c|ccc}
    & c_{\mathcal{OO}} & \gamma_{JJ \mathcal{O}} & \lambda^2_{JJ \mathcal{O}} \\
    \hline
    {\mathcal O}_{1,+} & 1 & 2 &  1 \\
    {\mathcal O}_{4,+} & 9 & 3 & \frac{1}{4} \\
    {\mathcal O}_{6,+} & 420 & 6 & \frac{3}{140} \\
    {\mathcal O}_{7,-} & 8640 &  144 & \frac{3}{5} 
    \end{array}
\end{equation}

\subsection*{Generalized Free Vector Field}

The GFVF $V^\mu(x)$ is conserved $\partial_\mu V^\mu=0$, and so the two-point function takes the form
\begin{equation}
    \langle V^\mu(x) V^\nu(0) \rangle =  I^{\mu\nu}(x) / |x|^4 .
\end{equation}
Correlators follow from Wick's Theorem. The low-lying primaries are
\begin{eqnarray}
\label{GFVFops}
{\mathcal O}_{4,+} &=& V^\mu V_\mu \ , \nonumber \\
{\mathcal O}_{6,+} &=& 7(\partial_\nu V^\mu) (\partial_\mu V^\nu) 
-8 (\partial_\nu V_\mu) (\partial^\nu V^\mu) + 6(\partial^2 V^\mu) V_\mu
\ , \nonumber \\
{\mathcal O}_{5,-} &=& i \epsilon^{\mu\nu\lambda} V_\mu \partial_\nu V_\lambda \ , \\
{\mathcal O}_{7,-} &=& i \epsilon^{\mu\nu\lambda} \left[ 
(\partial_\rho V_\mu) (\partial^\rho \partial_\nu V_\lambda) - 4 (\partial^2 V_\mu) ( \partial_\nu V_\lambda) 
- V_\mu (\partial^2 \partial_\nu V_\lambda)
\right] \nonumber \ .
\end{eqnarray}

\begin{equation}
    \begin{array}{c|ccc}
    & c_{\mathcal{OO}} & \gamma_{VV \mathcal{O}} &\lambda^2_{VV \mathcal{O}} \\
    \hline
    {\mathcal O}_{4,+} & 6 & 1 & \frac{1}{6} \\
    {\mathcal O}_{6,+} & 90{,}720  & 36 & \frac{1}{70} \\
    {\mathcal O}_{5,-} & 12 & 4 & \frac{4}{3} \\
    {\mathcal O}_{7,-} & 12{,}960 & 72 & \frac{2}{5} 
    \end{array}
\end{equation}

\subsection*{The Gross-Neveu-Yukawa Model}

The $O(N)$-symmetric GNY model can be described by the Lagrangian
\begin{equation}
    \mathcal{L}_{\rm GNY} = - \frac{1}{2} (\partial \phi)^2 - i \frac{1}{2}\psi_i \slashed \partial \psi_i - \frac{1}{2} m^2 \phi^2 - \frac{\lambda}{4}\phi^4 - i\frac{g}{2}\phi \psi_i \psi_i \ ,
\end{equation}
in $d=4-\epsilon$ dimensions. 
$\psi_i$ is $N$ Majorana fermions, which transform in the vector representation of $O(N)$. They interact with $\phi$, an $O(N)$-singlet pseudoscalar. At a critical value of $m^2$ and $\epsilon = 1$, we expect this theory to flow to a CFT where the only relevant even parity $O(N)$-singlet scalar operator is $\epsilon \sim \phi^2$. The lowest dimension odd parity scalar operator is $\sigma \sim \phi$. 
In \cite{Erramilli:2022kgp}, a bootstrap was performed in which the dimensions of $\epsilon$ and $\sigma$ were computed for $N=2,4$ and $8$. The values are
\begin{equation}
    \begin{array}{c|cc}
        N & \Delta_\sigma & \Delta_\epsilon \\
        \hline
        2 & 0.65 & 1.725 \\
        4 & 0.7578 & 1.899 \\
        8 & 0.8665 & 2.002
    \end{array}
\end{equation}

\subsection*{The $O(2)$ and $O(N)$ Models}

The $O(N)$ model is described by the Lagrangian 
\begin{equation}
    \mathcal{L}_{O(N)} = \frac{1}{2}(\partial \phi_i)^2 + \frac{1}{2}m^2 \phi_i^2 + \lambda (\phi_i^2)^2,
\end{equation}
again in $d=4-\epsilon$ dimensions,
where $\phi_i$ is $N$ scalars in the vector representation of $O(N)$. For tuned values of $m$ and $\lambda$, we expect to find a critical theory at $\epsilon =1$.  We are interested in the $N=2$ and large $N$ limit in particular. In these cases, the dimensions of the lowest-lying even parity scalar operators, denoted $\Delta_S$, are known
% \sam{Or did we get them from a more recent source?} 
to be
\begin{equation}
    \begin{array}{c|c}
        N & \Delta_S \\
        \hline
        2 & 1.51136 \\
        {\infty} & 2
    \end{array}
\end{equation}
The $O(2)$ result comes from  \cite{Kos_2014,Chester:2019ifh}, while the large $N$ result is a consequence of the Hubbard-Stratonovich trick and the fact that the theory is essentially free at leading order in $1/N$.

\subsection*{Odd Parity Dimension Five Scalars}

While the GFVF has the odd parity, dimension five operator ${\mathcal O}_{5,-} = i \epsilon^{\mu\nu\lambda} V_\mu \partial_\nu V_\lambda$ as discussed above, this operator is missing from the free fermion and complex scalar theories with a single $U(1)$ global symmetry \cite{Dymarsky:2017xzb}.  Naively we could try to build such an operator from the conserved current $V_\mu = J_\mu$ in each instance, but the construction will fail.  In the free scalar case where $J_\mu \sim \phi^* \partial_\mu \phi - \phi \partial_\mu \phi^*$, any term in the expansion of $J_\mu \partial_\nu J_\lambda$ will contain pieces completely symmetric in two of the indices, either $\partial_\mu \partial_\nu \phi$, $(\partial_\mu \phi) (\partial_\nu \phi)$ or their complex conjugates. 
% In the fermionic case, 
% by a Fierz identity, one can always reduce
% this dimension five operator in three dimensions to something that vanishes on-shell, up to a total derivative of the diment.
In the fermionic case, by a Fierz identity, this dimension five operator in three dimensions turns out to be a descendant of the dimension four operator $(\bar\psi \gamma^\mu \psi) (\bar\psi \psi)$.

If the global symmetry is $O(N)$ with $N \geq 4$, then we can
consider two mutually commuting currents,
corresponding to two mutually commuting $U(1)$ subgroups inside $O(N)$.  In this case, there is a dimension five operator 
${\mathcal O}_{5,-} = \epsilon^{\mu\nu\lambda} J^{(1)}_\mu \partial_\nu J^{(2)}_\lambda$.

\section{Details About the Numerics}\label{sec: numerics}

The implementation of our numerical bootstrap involved a pipeline of different code. The first piece was the pre-existsing software Blocks\_3d \cite{Erramilli:2020rlr}, which we fed into our own bespoke routines to realise the conservation conditions present in our theory. The result was further processed by our own routines in order to correctly format a numerical problem for the final piece of software, SDPB \cite{sdpb,Landry:2019qug}.

Blocks\_3d approximates three dimensional conformal blocks to arbitrary precision. The  representations and conformal dimensions of any external or exchanged operators can be completely arbitrary.  In a general numerical bootstrap, the conformal dimensions of external operators may need to be varied for each run of SDPB. However, in our case, the external currents are protected, with $\Delta = 2$. Unfortunately, this comes with its own complications: spinning operators have a more complicated tensor structure, and the currents obey a conservation condition. A limitation of Blocks\_3d is that the conformal blocks do not satisfy conservation conditions for external operators.

The second piece of our pipeline was to implement our own conservation routines onto the output of Blocks\_3d. These conservation rules involved taking linear combinations of various blocks, to produce what we called the conserved blocks. This non-trivial task was implemented in Mathematica, following procedures detailed in \cite{He:2023ewx}.

The conserved blocks then needed to be fed into SDPB. However, to examine the space of CFTs described in the text, we needed the ability to implement gap assumptions, input isolated operators, and also vary the norm and objective. To achieve this, we built a number of C programs and bash scripts that could construct the relevant structures that we were interested in. There were C routines to pack the norm and objective specified for a problem, followed by routines to pack the scalars and spin two blocks, which were the parts that were changed most frequently. Next, any isolated operators were packed by another C routine, before finally the higher spin operators were packed. Due to the approximations produced by Blocks\_3d, the conformal blocks are ratios of polynomials. In some cases (specifically, in even spin exchanged blocks with $J \geq 2$, and the odd parity $J=2$ block), there was a factor of $x/x$. We had another routine to remove this part, at the suggestion that it would improve the behaviour of SDPB.\footnote{Personal communication P.~Kravchuk.}

In general, these C routines were called as part of bash scripts. Some of the routines, such as setting the scalar gaps, were required to be called before each run of SDPB, whilst others, like generating the norm, were only needed once for a large number of runs. Thus, we were able to partition our routines to avoid unnecessary repeated computations, just changing the relevant part of the problem. 

\subsection{The Crossing Equations}

For completeness, we include the crossing equations that we satisfied in our bootstrap. The technicalities are identical to our previous work in \cite{Bartlett-Tisdall:2023ghh}, so we do not repeat everything. Here, $\langle q_1 q_2 q_3 q_4 \rangle$ denotes a $\mathbb{Z}_2 \times \mathbb{Z}_2$ invariant orbit of structures $[q_1 q_2 q_3 q_4]$ specified by $q_i \in \{-j_i, -j_i + 1, \hdots, j_i\} $, where $j_i = 1$ is the spin of external operators in the four-point function. $g^\pm(z, \bar{z})$ denotes the conformal blocks output by Blocks\_3d

One can consistently take $\langle 1111 \rangle$, $\langle 0110 \rangle$, $\langle 0101 \rangle$, 
$\langle 0011 \rangle$ and $\langle 0000 \rangle$ to be the five bulk blocks discussed in section \ref{sec: setup}.  
$\langle{-}1{-}111 \rangle$ and $\langle{-}1 11 {-}1\rangle$ are a convenient choice for the additional line blocks, which
furthermore exchange under
$1 \leftrightarrow 3$.  
The crossing equations are then
\begin{eqnarray}
\partial_z^n \partial_{\bar z}^{\bar n} g^+_{[1111]}(1/2, 1/2) &=& 0 \ , \; \; \; (n+\bar n \; \mbox{odd}) \ , \\
\partial_z^n \partial_{\bar z}^{\bar n} g^+_{[0101]}(1/2, 1/2) &=& 0 \ , \; \; \; (n+\bar n \; \mbox{odd}) \ , \\
\partial_z^n \partial_{\bar z}^{\bar n} g^+_{[0000]}(1/2, 1/2) &=& 0 \ , \; \; \; (n+\bar n \; \mbox{odd}) \ , \\
\partial_z^n \partial_{\bar z}^{\bar n} g^+_{[0011]}(1/2, 1/2) &=&  (-1)^{n+\bar n} 
\partial_z^n \partial_{\bar z}^{\bar n} g^+_{[1001]}(1/2, 1/2) \ .
\end{eqnarray}

\begin{eqnarray}
\partial_{\bar z}^n\partial_z^n g^+_{[-1-111]}(1/2, 1/2) &=&  \partial_{\bar z}^n \partial_z^n g^+_{[1-1-11]}(1/2, 1/2) \ , \\
\partial_{\bar z}^{n} \partial_z^{n+1} g^+_{[-1-111]}(1/2, 1/2) &=& 
 -\partial_{\bar z}^n \partial_z^{n+1} g^+_{[1-1-11]}(1/2, 1/2) \ .
\end{eqnarray}
Note that keeping purely holomorphic or antiholomorphic derivatives in the line constraint will fail to provide enough data to satisfy the conservation condition. Furthermore, we were not able to successfully implement a line constraint in SDPB in every case. We found that for $\Lambda > 18$, we were unable to increase the maximal $n$ in the line constraint beyond 18. There may be a reason that we have not understood that makes the higher order line equations degenerate; indeed, our crossing equations are different from \cite{He:2023ewx}, where they find a point constraint instead of a line constraint. However, we believe that it is related to the \texttt{kept\_pole\_order} parameter in Blocks\_3d, which specifies the number of poles to keep in the approximation of the conformal blocks. By experimenting with a lower \texttt{kept\_pole\_order}, we were able to reduce our maximal value of $n$ beyond which we were unable to satisfy crossing. Thus, we expect (although did not verify) that increasing \texttt{kept\_pole\_order} should allow us to satisfy the line constraint at higher order.

\subsection{Parameters Used for SDPB and Blocks\_3d}

\vspace*{1em}
\begin{center}
\begin{tabular}{|l|c|}
\hline
%maxIterations & 500 \\
dualityGapThreshold & 1e-30 (1e-6 for OPE bounds) \\
primalErrorThreshold & 1e-30 \\
dualErrorThreshold & 1e-20 \\
initialMatrixScalePrimal & 1e20 \\
initialMatrixScaleDual & 1e20 \\
feasibleCenteringParameter & 0.1 \\
infeasibleCenteringParameter & 0.3 \\
stepLengthReduction & 0.7 \\
maxComplementarity & 1e100 \\
\hline
\end{tabular}

\vskip 0.1in

\begin{tabular}{|l|c|c|c|c|}
\hline
$\Lambda$  & 17 & 18 & 25 \\
precision & 650 & 650 & 1024 \\
$\ell_{\rm max}$ & 30 & 30 & 50 \\
{\rm kept\_pole\_order}  & 20 & 20 & 20 \\
{\rm order} & 60 & 60 & 80 \\
\hline
\end{tabular}
\end{center}
\noindent
Despite some effort, we were never able to see dual jumps in the progress of SDPB, only primal jumps.  As a result, for the exclusion plots, an allowed point is one for which we observe a primal jump but a disallowed point corresponds to creeping behavior where both the dual and primal errors gradually creep down.  To distinguish these behaviors, we set the dualErrorThreshold higher than the primalErrorThreshold but still well below the values of primal error where the jumps were observed.

\bibliographystyle{jhep}
\bibliography{bib.bib}

\end{document}